\documentclass[journal=jpccck,manuscript=article,layout=twocolumn]{achemso}

\setkeys{acs}{abbreviations=true, articletitle=true,
biblabel=plain,
chaptertitle=true,
doi=true,
email=true,
etalmode=truncate,
keywords=true,
maxauthors=100,
super=true
}

\setcitestyle{super,open={},close={}}

\usepackage[utf8]{inputenc} 
\usepackage{textcomp} 
\usepackage[T1]{fontenc} 
\usepackage[portuguese,english]{babel} 
\usepackage[version=4]{mhchem} 
\usepackage{siunitx} 
\usepackage{graphicx} 
\graphicspath{{Figures/}} 
\usepackage{geometry} 
\usepackage[format=plain,justification=justified,singlelinecheck=false,font={normalsize,bf},labelfont=bf,labelsep=space]{caption} 
\usepackage{float} 
\usepackage{natbib} 
\usepackage{setspace} 
\usepackage{xkeyval} 
\usepackage{array} 
\usepackage{booktabs} 
\usepackage{listings} 
\usepackage{lmodern} 
\usepackage{mathpazo} 
\usepackage[breaklinks,colorlinks=true,allcolors=blue]{hyperref}
\hypersetup{colorlinks=true,citecolor=blue,urlcolor=blue}
\newcommand{\doi}[1]{\href{http://dx.doi.org/#1}{\nolinkurl{#1}}}

\usepackage{subfigure} 
\usepackage[compact]{titlesec} 
\usepackage{natmove} 
\usepackage{multirow} 
\usepackage{lipsum} 
\usepackage{calc} 
\usepackage{longtable} 
\usepackage{tabularx} 
\usepackage{placeins} 
\usepackage{tablefootnote}
\usepackage{adjustbox}
\usepackage{titletoc}
\usepackage{latexsym}
\usepackage{cancel}
\usepackage{amsmath}
\usepackage{amssymb}
\usepackage{amsfonts}
\usepackage{amstext}
\usepackage{float}

\DeclareSIUnit\angstrom{\text {Å}}
\DeclareSIUnit\bar{bar}

\usepackage[none]{hyphenat}
\usepackage{microtype}
\makeatletter
\let\l@addto@macro\relax
\makeatother
\usepackage[fontsize=12pt]{scrextend}

\SectionsOn
\SectionNumbersOn
\AbstractOn

\title[Title]{On the mechanical, thermoelectric, and excitonic properties of Tetragraphene monolayer}

\author{Raphael M. Tromer}
\affiliation[UNB]{State University of Campinas, Applied Physics Department , Campinas $13083-970$, SP, Brazil}

\author{L. A. Ribeiro Júnior}
\affiliation[UNB]{University of Bras{\'{i}}lia, Institute of Physics and International Center of Physics, Bras{\'{i}}lia $70919$-$970$, DF, Brazil}

\author{Douglas S. Galvão}
\affiliation[UNB]{State University of Campinas and Center for Computational Engineering and Sciences, Campinas $13083-970$, SP, Brazil}

\author{Alexandre C. Dias}
\affiliation[UNB]{University of Bras{\'{i}}lia, Institute of Physics and International Center of Physics, Bras{\'{i}}lia $70919$-$970$, DF, Brazil}

\author{Elie A. Moujaes}
\affiliation[UNIR]{Physics Department, Federal University of Rondônia, $76801-974$, Porto Velho, Brazil}
\email{eamoujaes@unir.br}

\begin{document}

\maketitle

\begin{tocentry}
\centering
\includegraphics[width=4.5cm,height=4.5cm]{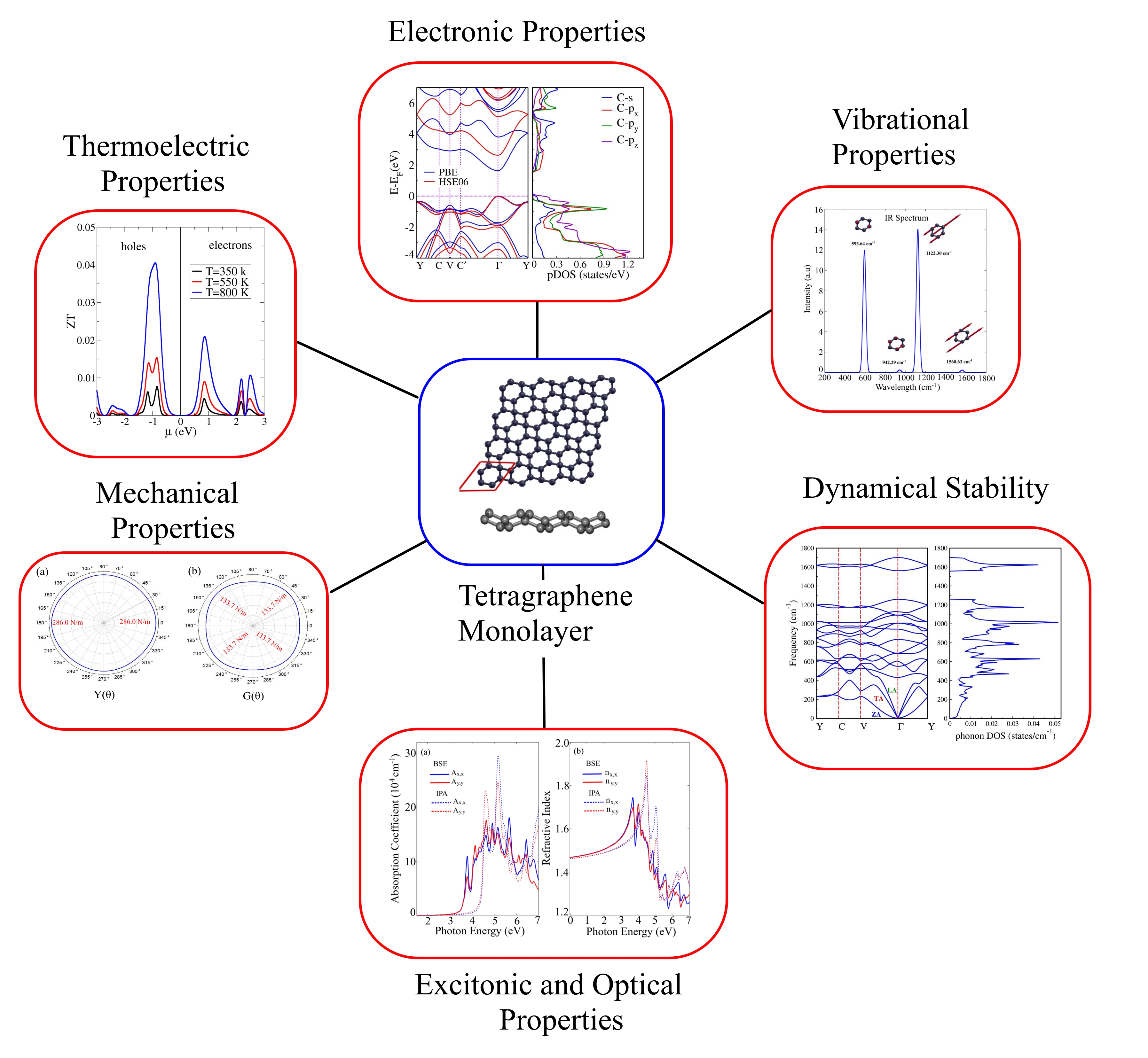}
\end{tocentry}

\begin{abstract}
Two-dimensional carbon allotropes have attracted much attention due to their extraordinary optoelectronic and mechanical properties, which can be exploited for energy conversion and storage applications. In this work, we use density functional theory simulations and semi-empirical methods to investigate the mechanical, thermoelectric, and excitonic properties of Tetrahexcarbon (also known as Tetragraphene). This quasi-2D carbon allotrope exhibits a combination of squared and hexagonal rings in a buckled shape.
~Our findings reveal that tetragraphene is a semiconductor material with a direct electronic bandgap of \SI{2.66}{\electronvolt}. Despite the direct nature of the electronic band structure, this material has an indirect exciton ground state of \SI{2.30}{\electronvolt}, which results in an exciton binding energy of \SI{0.36}{\electronvolt}. At ambient temperature, we obtain that the lattice thermal conductivity ($\kappa_L$) for tetragraphene is approximately 118 \si{\watt/\milli\kelvin}. Young's modulus and the shear modulus of tetragraphene are almost isotropic, with maximum values of \SI{286.0}{\newton/\meter} and \SI{133.7}{\newton/\meter}, respectively, while exhibiting a very low anisotropic Poisson ratio value of \SI{0.09}{}.
\end{abstract}

\section{Introduction}

Two-dimensional (2D) carbon-based materials have attracted considerable attention due to their extraordinary physical and chemical properties, which have been exploited in flat electronics applications.\cite{Novoselov2004,Novoselov2005} This fact is partly due to the revolution in materials science created by the advent of graphene.\cite{Novoselov2004,Novoselov2005}  Graphene has unique features combining high electrical conductivity, robust mechanical strength, and excellent thermal properties.\cite{geim2007rise,balandin2008superior}  These characteristics allow various applications, from organic electronics \cite{geim2007rise} to energy storage \cite{stoller2008graphene} and sensor technologies.\cite{geim2007rise, neto2009electronic}

In addition to graphene, many other 2D carbon structures, including graphyne,\cite{hu2022synthesis,desyatkin2022scalable,longuinhos2014theoretical} biphenylene monolayers,\cite{fan2021biphenylene,pereira2022mechanical} amorphous carbon,\cite{toh2020synthesis,felix2020mechanical} and fullerene networks,\cite{synth_full,junior2022thermal,tromer2022dft} were inspected, most of which have been successfully synthesized. Despite these accomplishments, the search for novel carbon allotropes continues with an emphasis on structures that could overcome some specific limitations associated with graphene, mainly its zero electronic bandgap, which limits its use in digital electronic devices. 

Computational methodologies for material design have become an important and reliable tool in research for various compelling reasons. Theoretical predictions of novel nanostructured materials can precede their synthesis by many years or even decades. Some examples include the biphenylene and the fullerene networks, which were theoretically predicted in 1987 \cite{baughman1987structure} and 1990,\cite{belavin2000stability} respectively. In this sense, the ability to simulate and predict material properties through computational models enables virtual experiments, thereby speeding up the exploration and advancement of novel materials. 

Recently, a novel quasi-2D carbon nanostructure named tetrahexcarbon was introduced by Ram and Mizuseki.\cite{ram2018tetrahexcarbon} This structure exhibits a combination of squared and hexagonal rings in a buckled shape (see Figure \ref{fig:crystal}). Due to this combination, an alternative name for the structure (tetragraphene \cite{de2019electronic}) has been proposed, and we adopted it in the present study. Tetragraphene contains carbon atoms in $sp^2$ and $sp^3$ hybridizations. It is a semiconductor material with a direct electronic bandgap of \SI{3.70}{\electronvolt}. It also presents high electronic mobility ($1.46\times 10^{-4}$cmV$^{-1}$s$^{-1}$) and relatively low cohesion energy.\cite{ram2018tetrahexcarbon}

Investigations into the electronic properties of tetragraphene-like structures have revealed metallic and semiconductor characteristics, with bandgap values ranging from \SI{0.82}{} up to \SI{4.4}{\electronvolt}, depending on structural parameters.\cite{de2019electronic} Studies on the mechanical properties of tetragraphene, using density functional theory (DFT) approaches, predicted values of Young's moduli along the zigzag and armchair directions of \SIrange{280}{288}{\newton/\meter}.\cite{ram2018tetrahexcarbon,de2019electronic,wei2020auxetic,kilic2020tuning} They also exhibit an intrinsic negative Poisson’s ratio and predicted rupture stresses and strains along zigzag and armchair directions of \SIrange{40}{50}{\percent} and \SIrange{33}{40}{\percent}, respectively.\cite{wei2020auxetic} Based on these unique characteristics, we comprehensively explored tetragraphene’s thermoelectric and excitonic properties to address its potential to serve as an active layer in thermal transport and energy conversion applications. To our knowledge, no studies have been reported on these relevant aspects of tetragraphene.

In this work, we use DFT simulations to investigate the mechanical, thermoelectric, and excitonic properties of tetragraphene. Through our computational approach, we obtained valuable information about this material's electronic and structural characteristics. In addition, we have also inspected optical and vibrational properties, along with a discussion of their Raman spectra.

\section{Theoretical Methods}

\subsection{Computational details}

The electronic band structure and density of states (DOS) were acquired using the plane wave-based Quantum Espresso (QE) code.\cite{QE} The structural geometry was optimized using a variable cell relaxation procedure until the force acting on each atom is less than \SI{e-4}{Ry/au} and the pressure is not larger than \SI{0.16}{\kilo\bar}. Scalar relativistic projector augmented wave (PAW) potentials \cite{blochl1994projector,kresse1999ultrasoft} within the Perdew-Burke-Ernzerhof (PBE) parametrization \cite{PBE} in the Generalized Gradient approximation (GGA) \cite{perdew1986accurate} were used for the exchange-correlation functional. A cutoff energy of \SI{60}{Ry} and a 34$\times$34$\times$1 dense electronic grid were sufficient to achieve a good convergence of the total energy value. A \SI{17}{\angstrom} vacuum buffer layer was added along the non-periodic $z$ direction to avoid undesirable interactions of the unit cell with its images. The electronic transport properties were obtained using the BOLTZTRAP code\cite{Madsen_67_2006} with an electronic \textbf{k}-mesh of 60$\times$60$\times$1.

To improve the electronic bandgap description, which is underestimated in the PBE framework due to self-interaction errors,\cite{Cohen_115123_2008,Crowley_1198_2016} we used the hybrid exchange-correlation functional proposed by Heyd--Scuseria--Ernzerhof (HSE06).\cite{heyd2004efficient,hummer2009heyd,moussa2012analysis} The Fock exchange was treated using the adaptive compressed exchange (ACE) method \cite{Lin_2242_2016} with a \textbf{k}-mesh of 6$\times$6$\times$1. Due to the heaviness of the calculation, we kept a 12$\times$12$\times$1 electronic \textbf{k}-mesh to deal with the PBE part of the Hamiltonian. Also, we used the same cutoff energy of the PBE simulation.

To analyze the structural stability of the tetragraphene (TG) monolayer, a phonon dispersion spectra over a 6$\times$6$\times$1 phonon (\textbf{q}) grid was also computed using a density function perturbation theory approach (DFPT),\cite{giannozzi2005density,baroni2010density} implemented in QE. A 24$\times$24$\times$1 \textbf{q}-grid was used to obtain well-converged phonon density of states (PhDOS). Using the QERAMAN code \cite{hung2024qeraman} interfaced with QE, first-order resonant Raman spectra at various laser energies were also estimated.

The elastic constants of the TG monolayer were determined using the THERMO$\_$PW code,\cite{malica2020quasi,malica2021quasi}, which is also interfaced with QE. The code generally applies strain along several directions of the material and fits several strain-stress curves to extract the corresponding elastic constants. For the case of the 2D TG, a total of twelve relaxation steps and an electronic 34$\times$ 34$\times$1 grid produced converged values. 

To probe the excitonic effects, we used a maximally localized Wannier functions tight binding (MLWF-TB) Hamiltonian to deal with quasi-particle effects and optical properties, as implemented in the WanTiBEXOS code.\cite{Dias_108636_2022} The MLWF-TB Hamiltonian was obtained at the HSE06 level from the Wannier90 package.\cite{mostofi2008wannier90} The Wannierization procedure was performed considering the $s$ and $p$ orbitals of the carbon atoms. The Bethe--Salpeter equation (BSE) \cite{Salpeter_1232_1951,Dias_3265_2021} was solved using a Coulomb truncated 2D potential (V2DT) \cite{Rozzi_205119_2006} and a \textbf{k}-mesh of 33$\times$33$\times$1. For the BSE solution, the highest three valence bands and the lowest two conduction bands were taken into account. A smearing of \SI{0.05}{\electronvolt} was utilized to compute the absorption coefficient at the BSE and independent particle approximation (IPA) levels.

\section{Results}

\subsection{Structural Properties}

\begin{figure}[!h]
    \centering
    \includegraphics[width=1\linewidth]{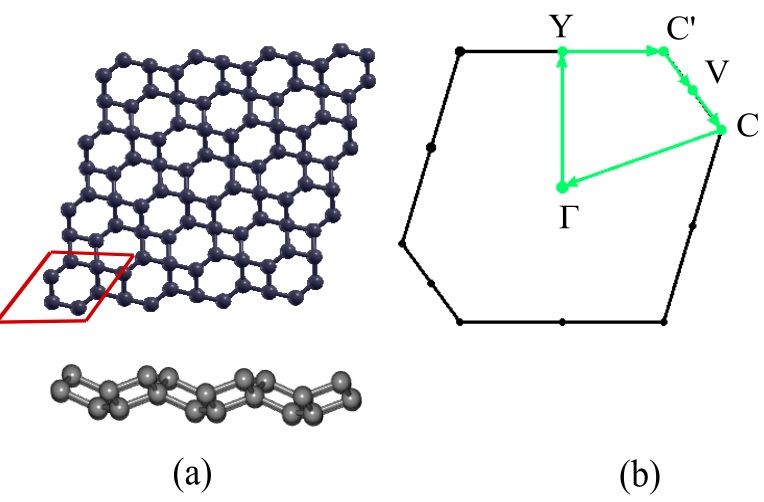}
    \caption{ (a) Top and side views of a TG monolayer. The red parallelogram indicates the primitive unit cell. (b) The first Brillouin zone and its respective high symmetry \textbf{k}-points.}\label{fig:crystal}
\end{figure}

The top and side views of the TG monolayer are shown in Fig~\ref{fig:crystal}(a), and the corresponding Brillouin zone (BZ) with the high symmetry points are displayed in Fig~\ref{fig:crystal}(b). The primitive unit cell contains six \ce{C} atoms and the structure belongs to the $Cmm2$ space group, with a lattice constant of \SI{3.79}{\angstrom} and an angle of \SI{73.19}{\degree} between the in-plane lattice vectors. Unlike graphene, the TG monolayer is not planar and has a buckling distance of \SI{0.58}{\angstrom} between two out-of-plane carbon atoms.

\begin{figure}[!h]
    \centering
    \includegraphics[width=1\linewidth]{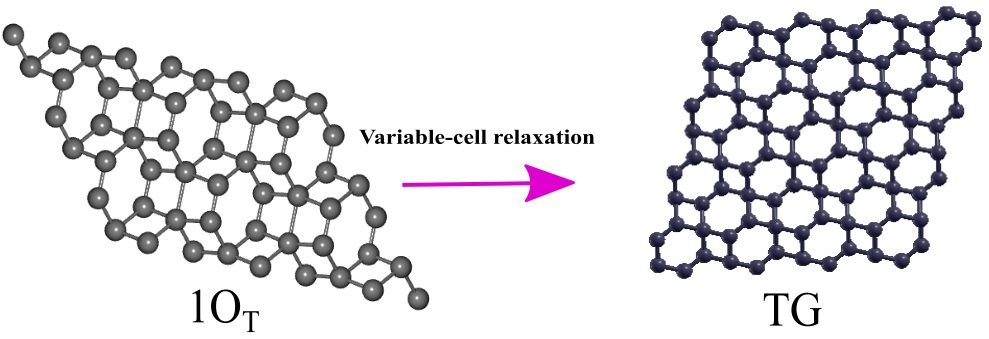}
    \caption{A variable cell relaxation simulation shows that a 1O$_{\rm{T}}$ carbon structure is unstable, transitioning to the more energy-favorable TG structure.}\label{fig:vc}
\end{figure}

Another new 2D structure with a unit cell containing six atoms is the recently predicted 1O$_{\rm{T}}$ structure \cite{moujaes2021optical} of transition metal dichalcogenides (TMDs), namely \ce{PdS2}, \ce{PdSe2}, and \ce{PdSSe}, whose excitonic properties have been recently investigated.\cite{moujaes2023excitonic} Its BZ is a tilted irregular hexagon, a signature of the monoclinic spacegroup (Fig. \ref{fig:vc}). We tested the carbon version of this topology using a variable cell relaxation simulation. We observed that the C-1O$_{\rm{T}}$ system is unstable and undergoes a phase transition to the TG structure (Fig. \ref{fig:vc}).  

The cohesive energy (E$_{\rm{coh}}$) of tetragraphene is determined by the values of its total energy and the energy of each carbon atom in the unit cell. In eV/atom:
\begin{equation*}
E_{\rm{coh}}=\frac{E_{TG}-6E_C}{6}~,  
\end{equation*}
where $E_{TG}$ is the total energy, and E$_C$ is the energy of an isolated C atom. We obtained the value of \SI{-8.36}{\electronvolt/atom}, indicating that the structure is energetically stable. This value is slightly larger than that of graphene (\SI{-9.23}{eV/atom} eV/atom) \cite{min2007ab} and graphyne (\SI{-8.582}{eV/atom} ),\cite{puigdollers2016first} but closer to the one for penta-graphene (\SI{-8.324}{eV/atom}).\cite{zhang2015penta}   


\begin{figure*}[!h]
    \centering
    \includegraphics[width=1\linewidth]{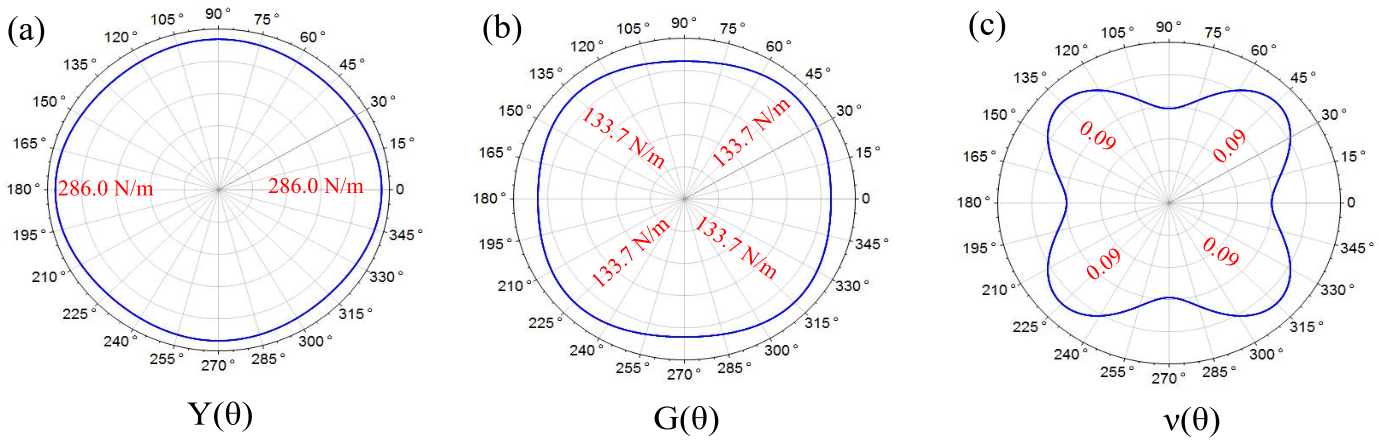}
    \caption{ 2D polar representation of (a) Young's modulus (N/m), (b) Shear modulus (N/m), and (c) Poisson's ratio in the $xy$ plane for the TG monolayer.}\label{fig:elastic}
\end{figure*}

\subsection{Elastic constants and mechanical stability}
The response of any material to the deformations under which it is subjected depends on several parameters, including its elastic constants. These constants are also important to estimate its mechanical stability. For a 2D system, in specific, the in-plane related elastic constants are the most important, namely C$_{ij}$ where \{i,j=1,2,6\}. For the TG monolayer, we obtained C$_{11}$=\SI{158.1}{\giga\pascal}, C$_{22}$=\SI{155.1}{\giga\pascal}, C$_{12}$=\SI{9.4}{\giga\pascal}, and C$_{66}$=\SI{69.4}{\giga\pascal}. Since C$_{11}$, C$_{22}$, C$_{66}$ > 0 and C$_{12}$ < C$_{11}$ and C$_{22}$, the material can be considered structurally stable. To obtain vacuum-independent results, the values have to be multiplied by 0.1$c$, where $c$ is the distance along the non-periodic $z$ direction. We further obtained C$_{11}$=\SI{287.1}{\newton/\meter} N/m, C$_{22}$=\SI{281.7}{\newton/\meter}, C$_{12}$=\SI{17.1}{\newton/\meter}, and C$_{66}$=\SI{126.0}{\newton/\meter}, which are near to the values reported in the work of Wei \textit{et al.}~.\cite{wei2020auxetic}

To verify  TG's structural anisotropy, orientation-dependent parameters should be determined. In particular, we evaluated Young's modulus (Y), the shear modulus (G), and the Poisson ratio ($\nu$) as a function of an in-plane angle $\theta$ (0$\leq\theta\leq 2\pi$), and the different compliance's S$_{ij}$=C$_{ij}^{-1}$ \{i,j=1,2,6\}\cite{jasiukiewicz2008auxetic,liu2021highly}:
\begin{eqnarray}
\frac{1}{{\rm{Y}}(\theta)}&=&S_{11}\cos^4{\theta}+S_{22}\sin^4{\theta} \nonumber \\
&+&2S_{16}\cos^3{\theta}\sin{\theta} \nonumber \\
&+&2S_{26}\cos{\theta}\sin^3{\theta}\nonumber \\
&+&(2S_{12}+S_{66})\cos^2{\theta}\sin^2{\theta}. \nonumber \\
\end{eqnarray}

\begin{equation}
 \nu(\theta)=-{\rm{Y}}(\theta) (A+B\psi).
\end{equation}

\begin{equation}
\frac{1}{4{\rm{G}}(\theta)}=C+D\cos{(4\phi+2)}.    
\end{equation}

\begin{equation}
A=\Big[\Big(S_{11}+S_{22}-S_{66}\Big)/2+3S_{12}\Big]/4,
\end{equation}

\begin{equation}
B=\frac{\sqrt{\Big(S_{26}-S_{16}\Big)^2+\Big(S_{12}-\frac{\Big(S_{11}+S_{22}-S_{66}\Big)}{2}\Big)^2}}{4},
\end{equation}

\begin{equation}
C=\frac{S_{11}+S_{22}-2S_{12}+S_{66}}{8},    
\end{equation}

\begin{equation}
D=\frac{\sqrt{\Big(S_{26}-S_{16}\Big)^2+\frac{\Big(S_{66}+2S_{12}-S_{11}-S_{22}\Big)^2}{4}}}{4}.
\end{equation}

$\psi$ and $\phi$ are defined in a way that:
\begin{equation}
\tan{\psi}=\frac{S_{26}-S_{16}}{S_{12}-\frac{\Big(S_{11}+S_{22}-S_{66}\Big)}{2}}  
\end{equation}

\begin{equation}
\tan{\phi}=\frac{2\Big(S_{16}-S_{26}\Big)}{S_{66}+2S_{12}-S_{11}-S_{22}}~.    
\end{equation}

From Fig. \ref{fig:elastic}, we realize that both Y and G are almost isotropic, with maximum values of \SI{286.0}{\newton/\meter} and \SI{133.7}{\newton/\meter}, respectively. However, the polar graph of the Poisson's ratio ($\nu$) shows an anisotropic behavior reaching a maximum value of 0.09 for $\theta$=45$^{o}$, 135$^{o}$, 225$^{o}$, and 315$^{o}$. Having $\nu$ < 0.5 satisfied further evidences the structural stability of the TG monolayer.

\subsection{Electronic Properties}


\begin{figure}[!h]
    \centering
    \includegraphics[width=1\linewidth]{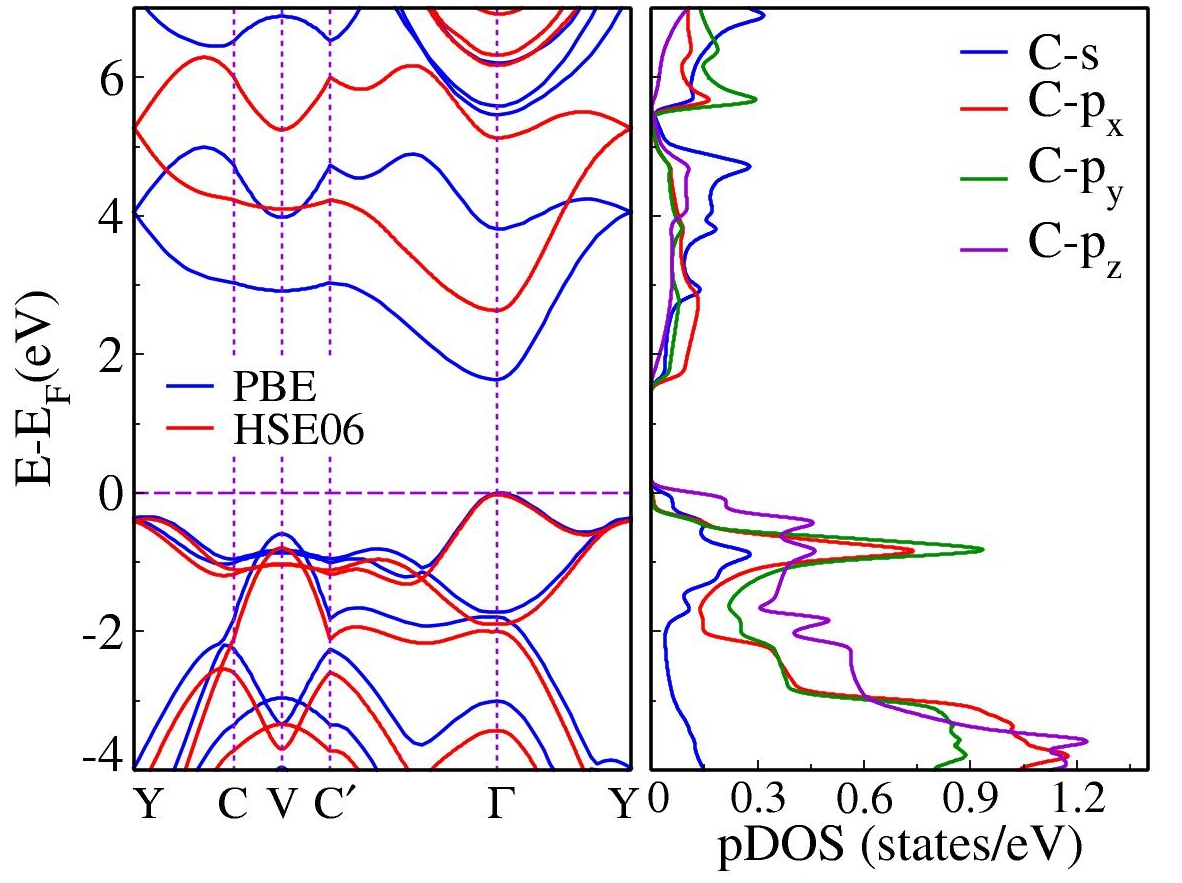}
    \caption{ (Left panel) Tetragraphene monolayer electronic band structure at the PBE (blue curves) and HSE06 (red curves) levels. (Right panel) Projected Density of States (PDOS) onto the carbon's $s$ and $p$ orbitals. The Fermi level value is set at \SI{0}{\electronvolt}. }\label{fig:bands_dos}
\end{figure}

The electronic band structure in the left panel of Fig.~\ref{fig:bands_dos} shows that the TG monolayer has a PBE(HSE06) direct band gap of \SI{1.63}{\electronvolt}(\SI{2.66}{\electronvolt}) localized at the high symmetry $\Gamma$ point. The projected density of states (PDOS) at the PBE level, shown in the right panel of Fig.~\ref{fig:bands_dos}, demonstrates that the top valence band (TVB) has a majority contribution from the $p_{z}$ orbital followed by a contribution from the $s$ orbital. However, the main contribution of the bottom conduction band (BCB) comes from the $p_{x}$ orbital, but also with contribution from the $s$ and $p_{y}$ orbitals. These features differ from those observed in graphene, where the $p_z$ orbitals majorly compose these states. This difference can be attributed to the buckling in the tetragraphene case, which additionally has a $\sigma$-bond character in the carbon-carbon interactions, in contrast to the $\pi$ electrons contributing to the BCB and TVB in graphene.

\subsection{Vibrational properties and Raman spectra}

\begin{figure}[!h]
    \centering
    \includegraphics[width=1\linewidth]{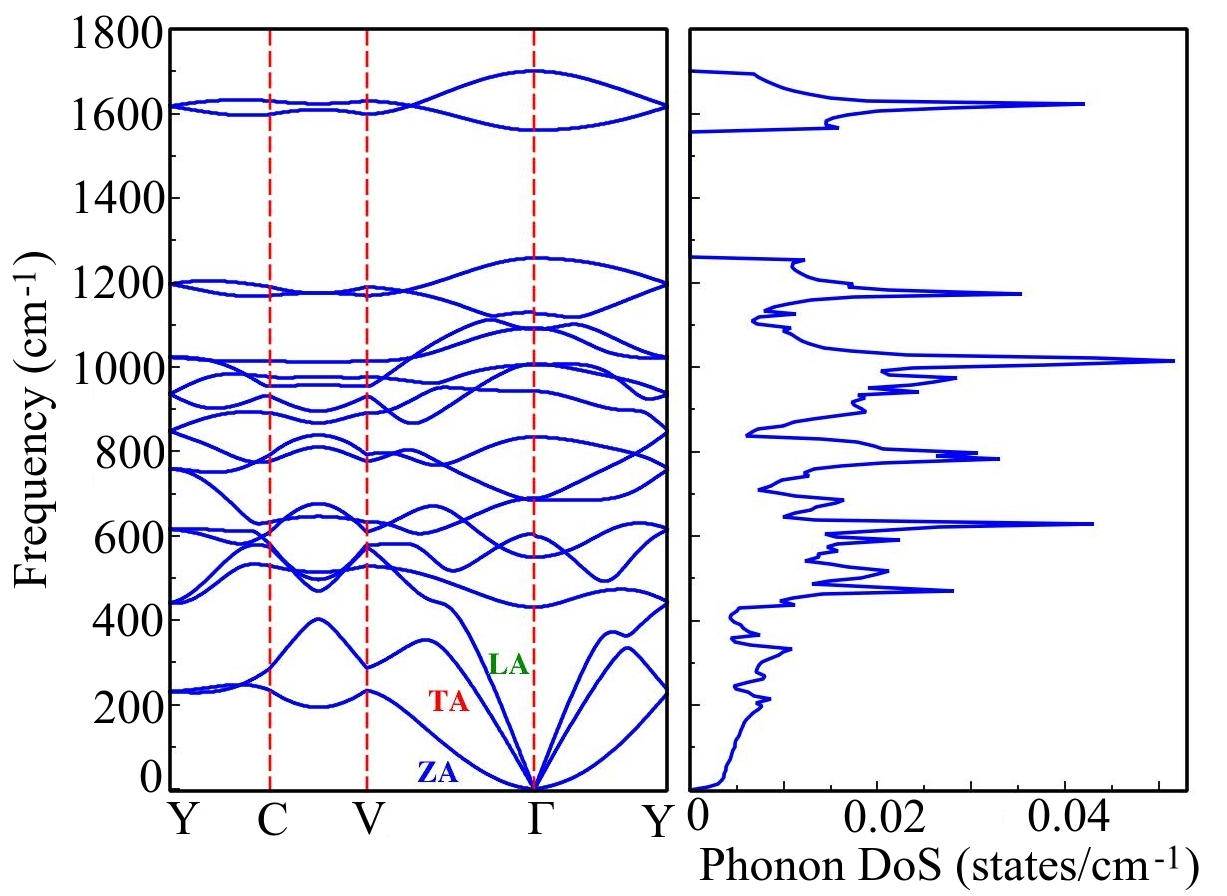}
    \caption{Phonon dispersion (left panel) of the TG monolayer and its phonon DOS (right panel).}\label{fig:phonon_dos}
\end{figure}

The phonon dispersion of the TG monolayer along the $\Gamma$-C-Y-V-$\Gamma$-Y symmetry path is displayed in Fig.~\ref{fig:phonon_dos}. Since the structure has six carbon atoms in its primitive unit cell, we expect eighteen vibrational modes, three of which are acoustic and the other fifteen optical. No negative phonon frequencies were observed along the selected path, consistent with a structurally stable structure. The flexural mode (ZA), the lowest phonon branch, follows a parabolic dispersion in the vicinity of the $\Gamma$ point, a signature of 2D materials, \cite{molina2011phonons,han2019tunable,huang2014correlation,qin2015anisotropic,peng2016towards,shao2018first} including graphene.\cite{jiang2015review} Conversely, the longitudinal (LA) and transverse (TA) acoustic modes show a linear dispersion behavior. 

Two interesting facts can be inferred from the phonon dispersion branches. Firstly, an optical-optical phonon bandgap of $\sim$ \SI{296}{\per\cm} can be observed between the low and high optical modes. The ability to tune such a bandgap by applying strain or stacking up more layers can be an essential feature in applications involving selective frequency transmission. Secondly, the highest four optical modes are flatter (low dispersion) than the others, translating into sharp peaks in the material's phonon DOS (PhDOS). Consequently, such modes have a small group velocity and, thus, do not significantly affect the thermal phonon transport properties.

\begin{figure*}[!h]
    \centering
    \includegraphics[width=1\linewidth]{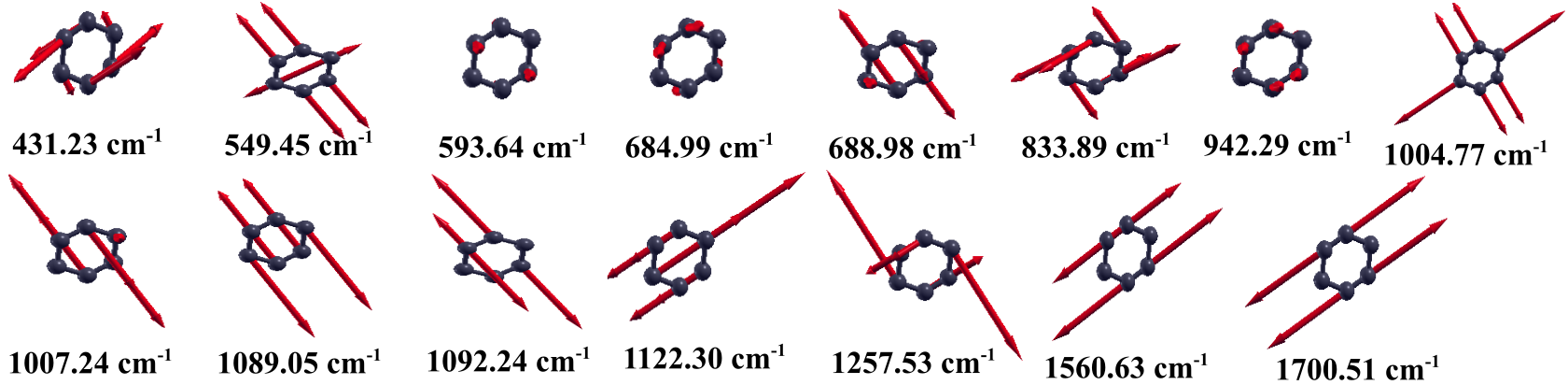}
    \caption{Optical phonon modes, at the $\Gamma$ point, of a TG monolayer. The arrows indicate the directions of the atoms' displacements.}\label{fig:ph-modes}
\end{figure*}

Fig.~\ref{fig:ph-modes} exhibits the fifteen optical modes of the TG monolayer. Optical modes 3 and 4 are characterized by almost perfect perpendicular vibrations of four of the six carbon atoms, while all atoms vibrate perpendicularly in the seventh mode. Nevertheless, most of the remaining modes have their carbon atoms vibrating within the plane of the structure, with small motions along the perpendicular direction, except for modes 5, 6, and 9.

\begin{figure*}[!h]
    \centering
    \includegraphics[width=1\linewidth]{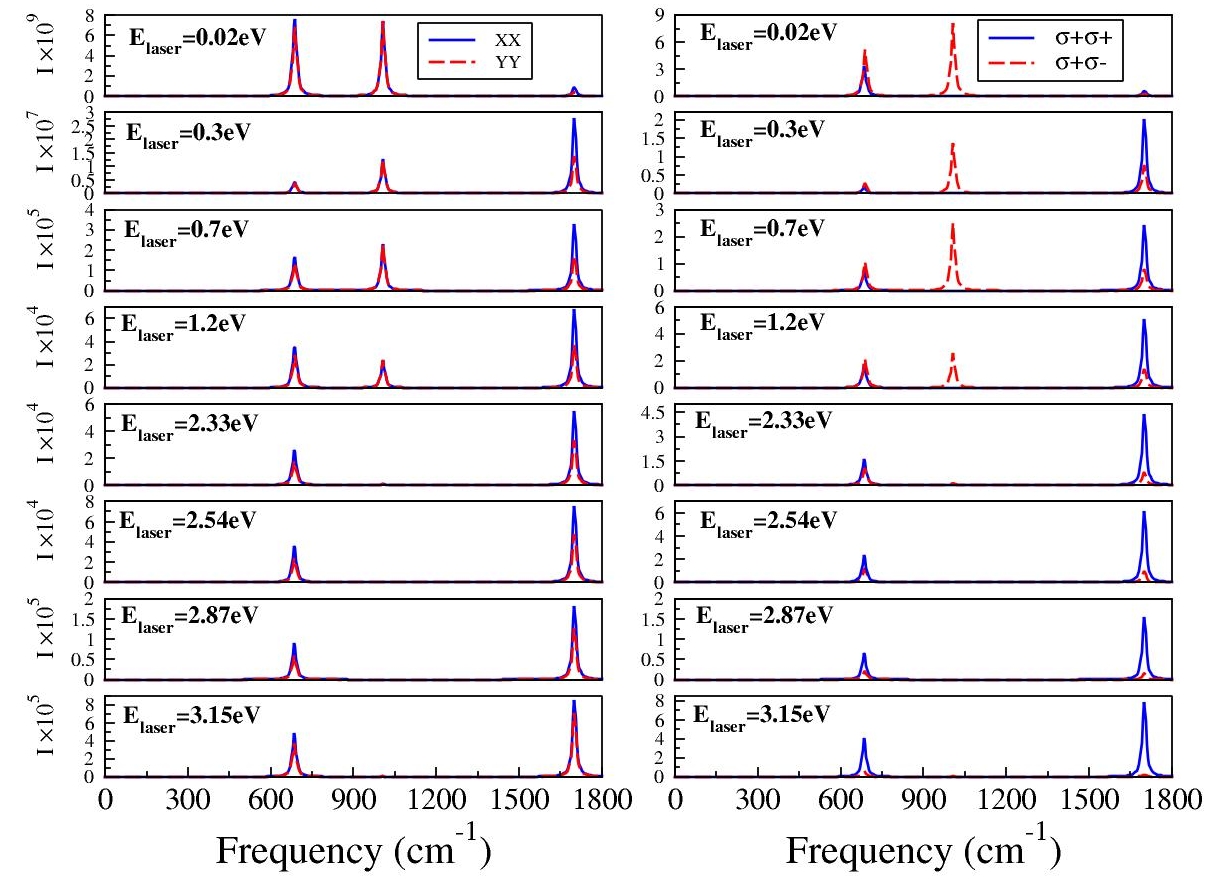}
    \caption{Raman spectra with laser excitation energies in the infrared, visible, and ultraviolet regimes for linearly (XX, XY) and circularly ($\sigma$+$\sigma$+, $\sigma$+$\sigma$-) polarized light. ``I'' refers to the intensity of the peaks in arbitrary units.}\label{fig:raman}
\end{figure*}

Not all phonon modes are Raman (R) or Infrared active (IR). Fig.~\ref{fig:raman} displays the calculated Raman spectra for linear and circular polarizations and at different laser energy values (E$_{\rm{laser}}$), ranging from the far IR to the ultraviolet (UV) regime. Three peaks occur at \SI{688.46}{\per\cm}, \SI{1007.6}{\per\cm}, and \SI{1700.58}{\per\cm}. We realize that the first and third peaks remain intense enough for the different laser energies. Yet, for values larger than or equal to \SI{2.33}{\electronvolt} and as E$_{\rm{laser}}$ increases, the second peak's intensity fades away, becoming negligible compared to the others. The first peak could be attributed to the fifth mode in Fig.~\ref{fig:ph-modes}, while the second peak could be ascribed to the ninth mode. The third peak corresponds to the fifteenth optical mode.

\begin{figure}[!htbp]
    \centering
    \includegraphics[width=1\linewidth]{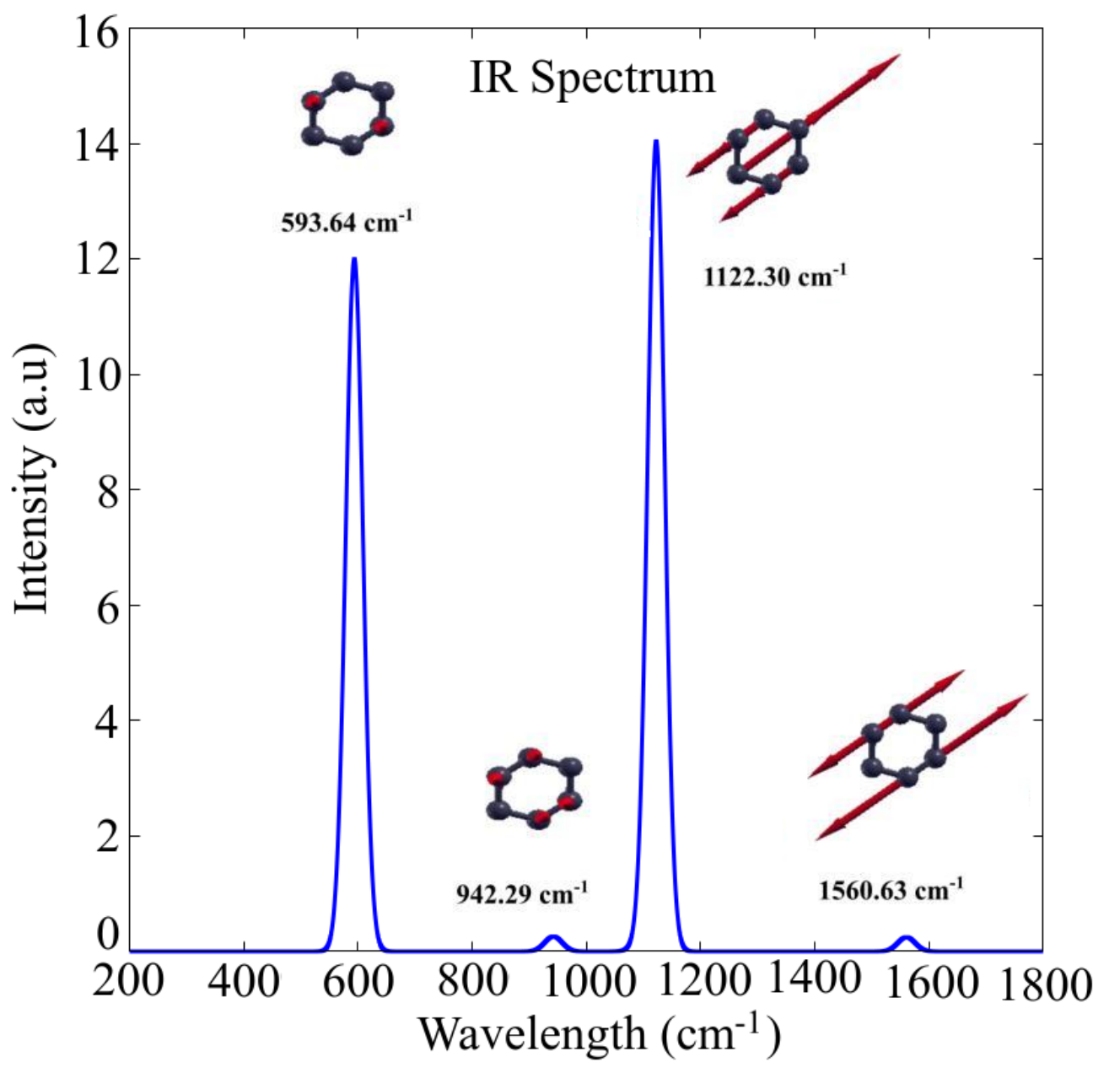}
    \caption{The IR spectrum of the TG monolayer and its corresponding vibrational modes.}\label{fig:ir}
\end{figure}

Naively, we might consider that all modes with zero intensity in the IR are Raman active. To ensure that only the modes mentioned above are Raman active, we calculated the intensity of the IR modes (Fig. \ref{fig:ir}). Our calculations reveal that optical modes 3, 7, 12, and 14 are IR-active.  We thus speculate that the remaining eight modes are silent; they are neither IR nor Raman active.



\subsection{Thermoelectric Properties}

\begin{figure}[!hb]
    \centering
    \includegraphics[width=1\linewidth]{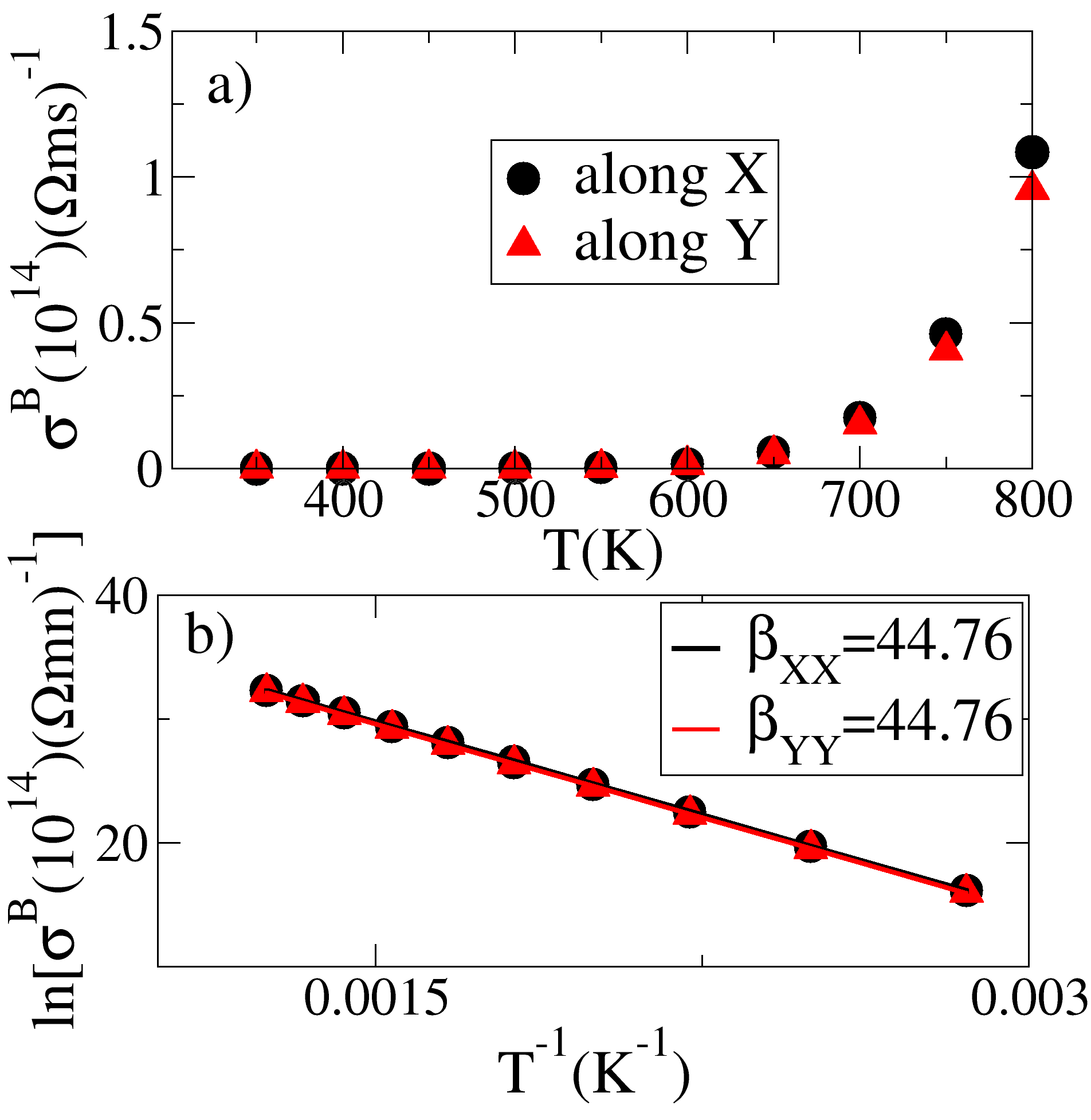}
    \caption{(a) Electrical conductivity tensors as a function of temperature, considering the X and Y directions within the 2D plane. (b) The Arrhenius plot required to determine the relaxation time, as described in reference \cite{Tromer2022}.}
    \label{fig:sigma_mu0}
\end{figure}

In Fig.~\ref{fig:sigma_mu0}-(a), we present the electrical conductivity tensor $\sigma^B$ calculated using the Boltztrap code without carrier excess, with a null chemical potential ($\mu=0$), as a function of temperature, considering the X and Y directions within the 2D plane. The electrical conductivity is almost zero for temperatures below \SI{600}{\kelvin}. This behavior is characteristic of semiconductors, as electrons would acquire sufficient energy to overcome the energy gap barrier only above a specific temperature threshold . We also observe that the electrical conductivity along both directions within the plane is nearly identical, with a slight difference at higher temperatures. Hence, as far as electrical transport is concerned, the system is essentially isotropic.

\begin{figure}[!htbp]
    \centering
    \includegraphics[width=1\linewidth]{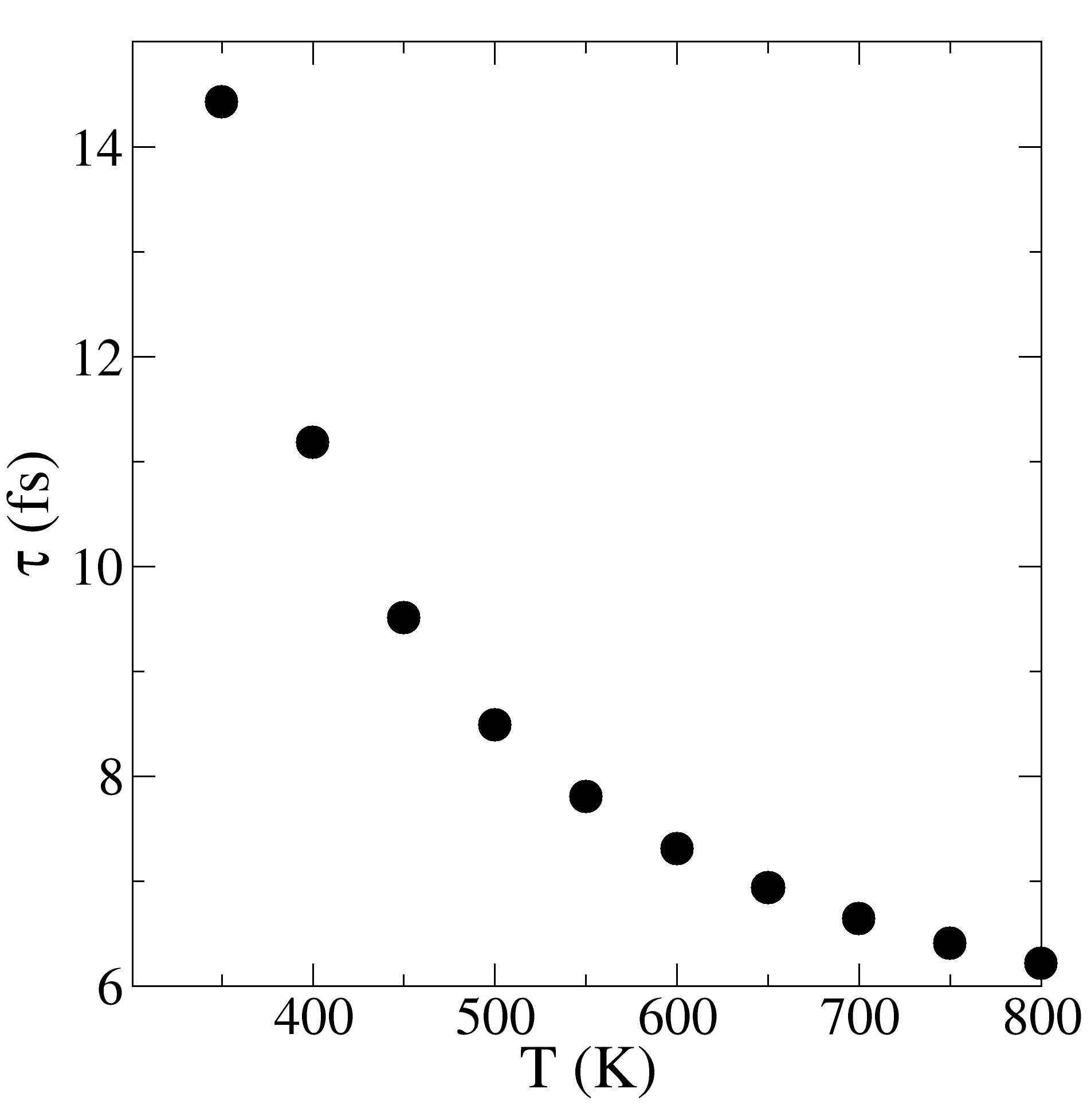}
    \caption{Electronic relaxation time ($\tau$) as a function of temperature of the TG monolayer.}
    \label{fig:tau}
\end{figure}

Fig.~\ref{fig:sigma_mu0}-(b) demonstrates the Arrhenius plot considering the electrical conductivity presented in Fig.~\ref{fig:sigma_mu0}-(a). From the plot's slope, we have estimated that tetragraphene's bandgap is approximately \SI{1.7}{\electronvolt}, a value very close to that deduced from the electronic band structure.  Constructing the Arrhenius plot enables us to determine the linear coefficient, discussed in the work of Tromer \textit{et al.} \cite{Tromer2022}, needed to estimate the electronic relaxation time $\tau$ as: 

\begin{equation}
    \tau (T) =\frac{\Gamma (T_0)}{\Gamma (T)}\frac{ln(\Gamma (T_0))}{ln(\Gamma^{Ref} (T_0))}\tau^{Ref.} (T_0),\label{eqn_t1}
\end{equation}
where $\tau^{Ref}$ is a previously calculated reference value for nitrogen-doped graphene at the reference temperature $T_0=350$ \si{\kelvin}. We will use the values $\tau^{Ref.}(350)=14.3$ \si{\fs} and, $\Gamma^{Ref.}(350)=11.997$, from Ref. \cite{Tromer2022}. The quantity $\Gamma$ is an auxiliary function given by:

\begin{equation}
    \Gamma (T)=ln \bigg[\frac{\sigma^{B}(T)}{\beta} \bigg].\label{eqn_t2}
\end{equation}
Once we have confirmed the system is isotropic, we will consider the average quantities along the 2D plane, where $\sigma=\frac{(\sigma^B_{XX}+\sigma^B_{YY})}{2}$ and $\beta=\beta_{XX}=\beta_{YY}$ is an estimated parameter derived from the BoltzTraP code. Fig.~\ref{fig:tau} shows the electronic relaxation time ($\tau$) as a function of temperature (T), using the scheme discussed previously. We note that $\tau(\rm{T})$ presents an exponential decay behavior, as expected for 2D systems.

\begin{figure}[!htbp]
    \centering
    \includegraphics[width=1.1\linewidth]{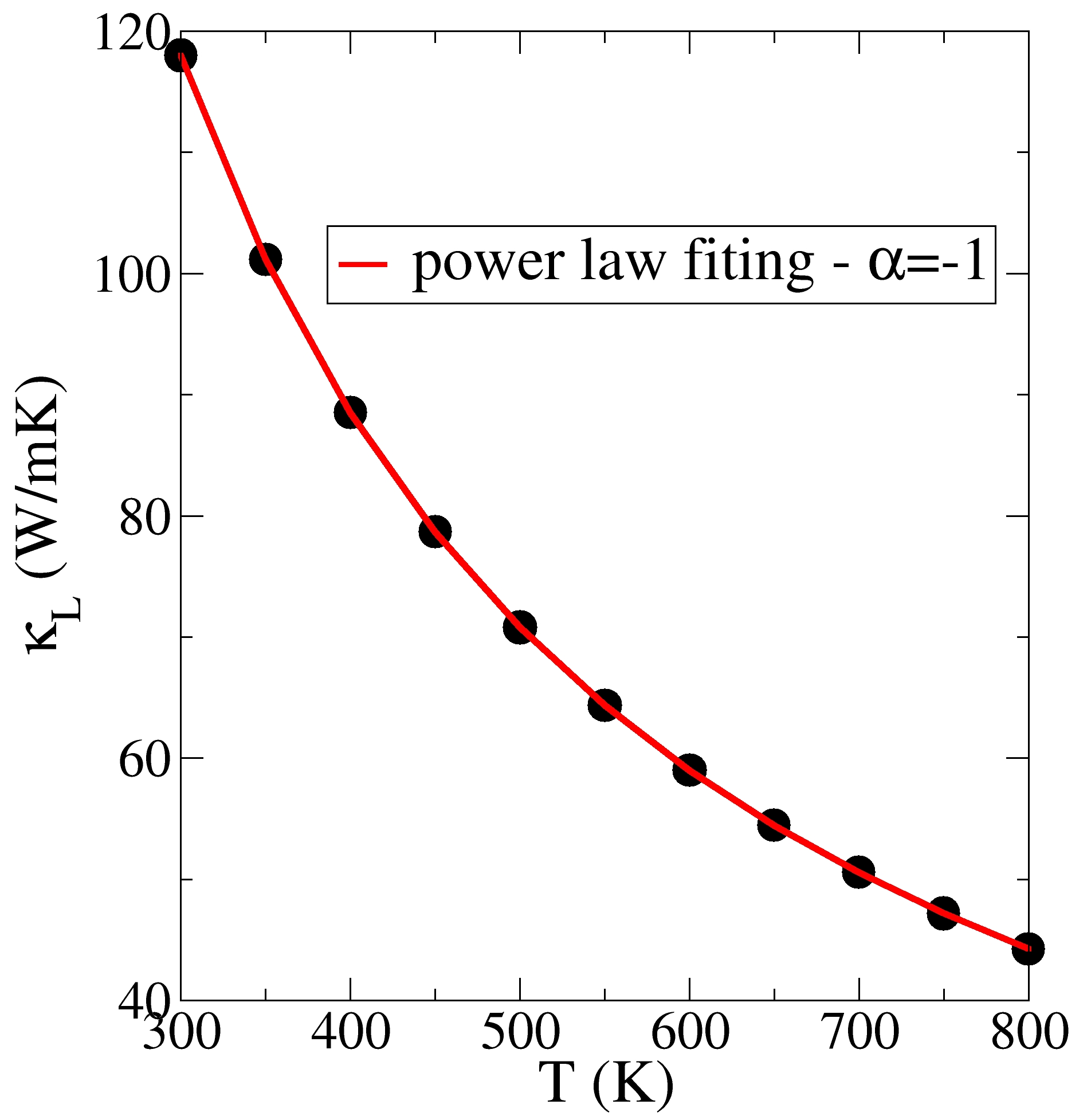}
    \caption{Lattice thermal conductivity $\kappa_L$ of the TG monolayer as a function of temperature.}
    \label{fig:kappa}
\end{figure}

Another critical parameter required for analyzing the thermoelectric properties of a particular system is the lattice thermal conductivity ($\kappa_L$). Using the method outlined in Ref.\cite{Tromer_28703_2023}, we estimate the value for tetragraphene to be approximately $\kappa_L(300)=118$ \si{\watt/\milli\kelvin}. This value is significantly lower than that of graphene ($3084.6$ \si{\watt/\milli\kelvin}), graphenylene ($600.0$ \si{\watt/\milli\kelvin}), and T-graphene ($800.0$ \si{\watt/\milli\kelvin}), but comparable to the thermal conductivity of the biphenylene network ($208.0$ \si{\watt/\milli\kelvin}) \cite{Tromer2022}.

To determine the values of $\kappa_L$ at higher temperatures, we assume that its temperature dependence follows a power-law relationship with an exponent equal to \num{-1}, a characteristic of 2D systems.\cite{Pereira2021} We will ,therefore, use a straightforward heuristic approach by defining the function as $118 \times$ ($300$/T), with T ranging from $300$ up to $800$. The values of $\kappa_L$ for different temperatures are depicted in Fig.~\ref{fig:kappa}. 

\begin{figure}[!ht]
    \centering
    \includegraphics[width=1\linewidth]{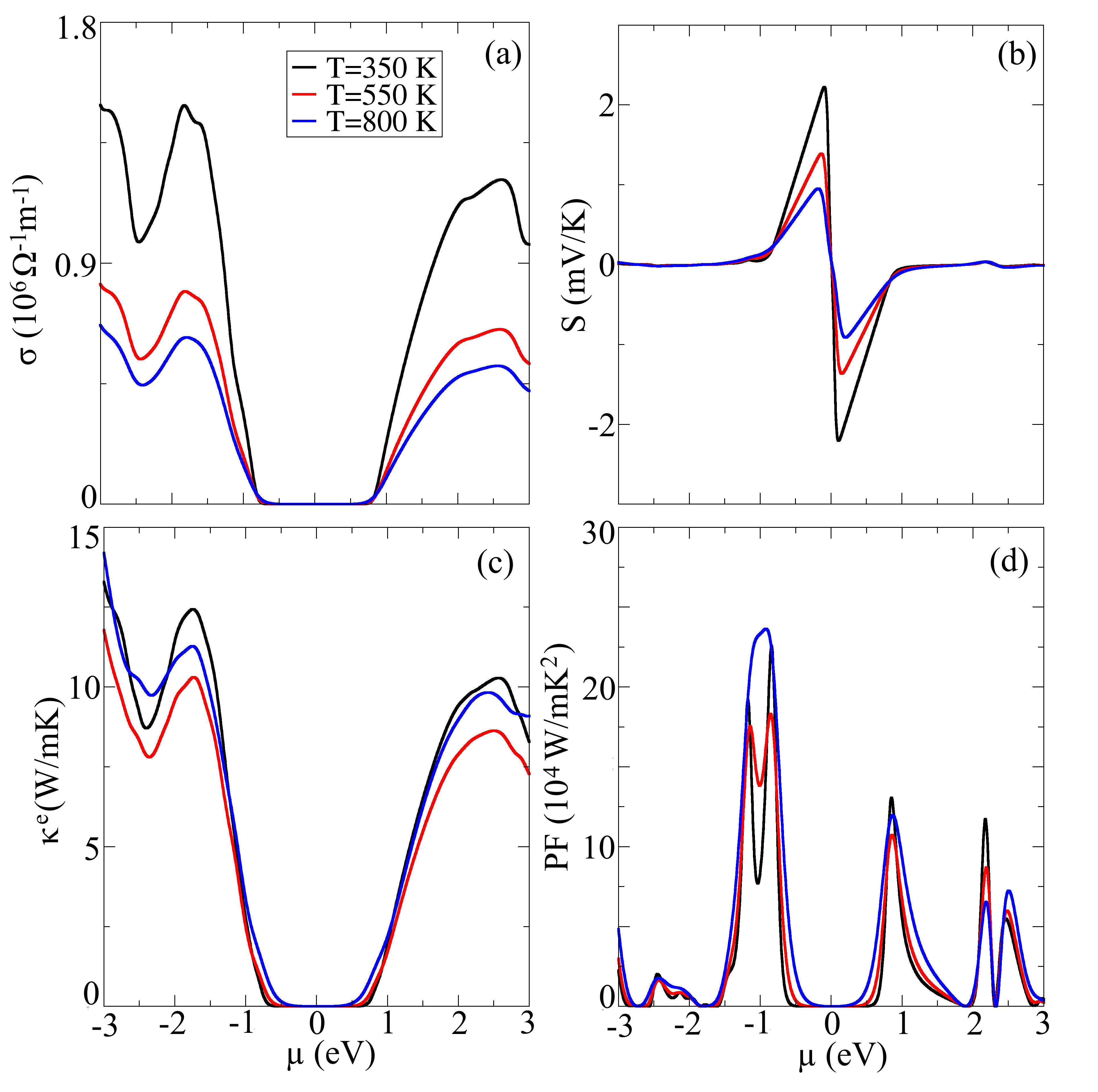}
    \caption{Thermoelectric coefficients as a function of positive (electron doping) and negative (hole doping) values of the chemical potential at different temperatures.}
    \label{fig:thermoelectric}
\end{figure}

Fig.~\ref{fig:thermoelectric} depicts the various thermoelectric properties as a function of a varying chemical potential $\mu$ (Fermi level at zero) at \SI{350}{\kelvin}, \SI{550}{\kelvin}, and \SI{800}{\kelvin}. Negative chemical potential values correspond to hole doping, while positive values denote electron doping. It is important to note that the electrical conductivity $\sigma$, the electronic contribution to thermal conductivity $\kappa^e$, and consequently, the power factor $PF$ ($S^2\times\sigma$) have been adjusted by the relaxation time values.

In Fig.~\ref{fig:thermoelectric}-(a), we observe a notable null region near the Fermi level for the electrical conductivity. This region corresponds to the energy gap of the material, where electrons lack sufficient energy to transition from the valence band to the conduction band. Only for slightly higher chemical potential values does the electrical conductivity exhibit non-zero values due to the formation of excitons. Additionally, the material displays more efficient hole transport than electron transport, maximizing $\sigma$ in the case of holes. Furthermore, the electrical conductivity decreases with increasing temperature, which is attributed to the fact that at higher temperatures, the number of charge carriers increases, affecting the mobility and reducing the electrical conductivity.

The Seebeck coefficient $S$ is presented in Fig.~\ref{fig:thermoelectric}-(b), which physically represents the magnitude of the thermoelectric voltage generated in response to a given temperature. Notably, the figure shows a nearly symmetrical behavior with respect to $\mu=0$. Moreover, the highest contribution to the Seebeck coefficient occurs in regions around the Fermi level.

The contribution of electrons/holes to the material's thermal electronic conductivity is shown in Fig.~\ref{fig:thermoelectric}-(c). Compared to the lattice thermal conductivity shown in Fig.~\ref{fig:kappa}, we observe that the contribution of electrons/holes is at least one order of magnitude lower. Hence, phonons play a substantially significant role in the material's thermal conductivity. To maximize the material's thermoelectric efficiency, exerting control over phonons is paramount.

A critical parameter in the context of thermoelectric materials and devices, which significantly influences their performance, is the power factor ($PF=S^2 \times \sigma$), presented in Fig.~\ref{fig:thermoelectric}-(d). For the TG monolayer, we observe that the power factor reveals a peak efficiency when the chemical potential approaches approximately $\pm1$\si{\volt} due to the distinct contributions from holes and electrons. Notably, the maximum power factor occurs at different temperatures for holes and electrons, with \SI{350}{\kelvin} and \SI{800}{\kelvin} being the optimal temperatures. These findings indicate the critical influence of the chemical potential and temperature on thermoelectric performance, highlighting the importance of carefully considering these factors in designing and optimizing thermoelectric materials and devices.

The power factor provides valuable insights into a material's suitability for thermoelectric applications by highlighting its ability to convert thermal gradients into electrical power. However, it is important to note that while a high $PF$ is desirable, it does not provide a complete picture of a material's overall thermoelectric efficiency. Other factors, such as thermal conductivity and device design, must also be considered to fully determine the performance of thermoelectric systems.

The quantity that comprehensively determines the thermoelectric conversion efficiency is known as the figure of merit, denoted as ZT, and is evaluated as follows: 

\begin{equation}
    ZT=\frac{S^2 \sigma}{(\kappa^e+\kappa_L)}T.\label{eqn_t3}
\end{equation}
Note that $ZT$ is the ratio of the power factor quantity and the thermal conductivity multiplied by the temperature parameter $T$.

\begin{figure}[!ht]
    \centering
    \includegraphics[width=1\linewidth]{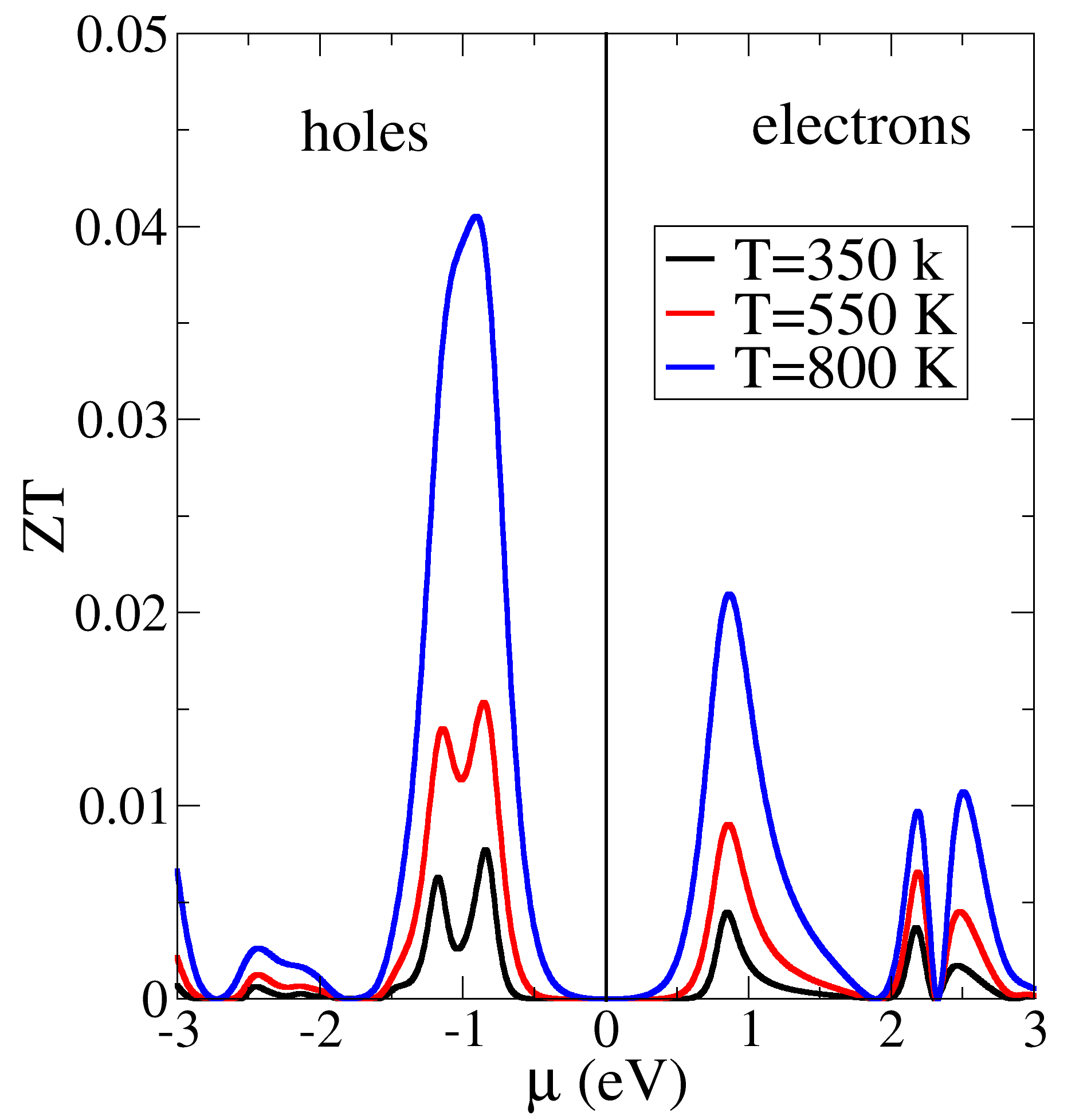}
    \caption{Figure of merit ZT as a function of the chemical potential for different temperatures $T$.}
    \label{fig:ZT}
\end{figure}

Finally, Fig.~\ref{fig:ZT} portrays the variation of ZT with the chemical potential at three disparate temperatures for the TG monolayer. Akin to the case of the PF, the peak values are also located at chemical potential values of approximately $\pm 1$. These values are highly temperature-dependent and tend to increase with increasing temperature. This behavior is due to the substantial decrease in the thermal conductivity of the lattice $\kappa_L$ with increasing temperature, as observed in Fig.~\ref{fig:kappa}. 
The maximum efficiency for holes ($0.04$) is nearly twice that obtained for electrons ($0.02$) at $T=800$ \si{\kelvin}. At lower $T$ values, the peak intensities are similar. Still, those associated with holes are always slightly larger.

\subsection{Excitonic and Optical properties}


\begin{figure}[!h]
    \centering
    \includegraphics[width=1.1\linewidth]{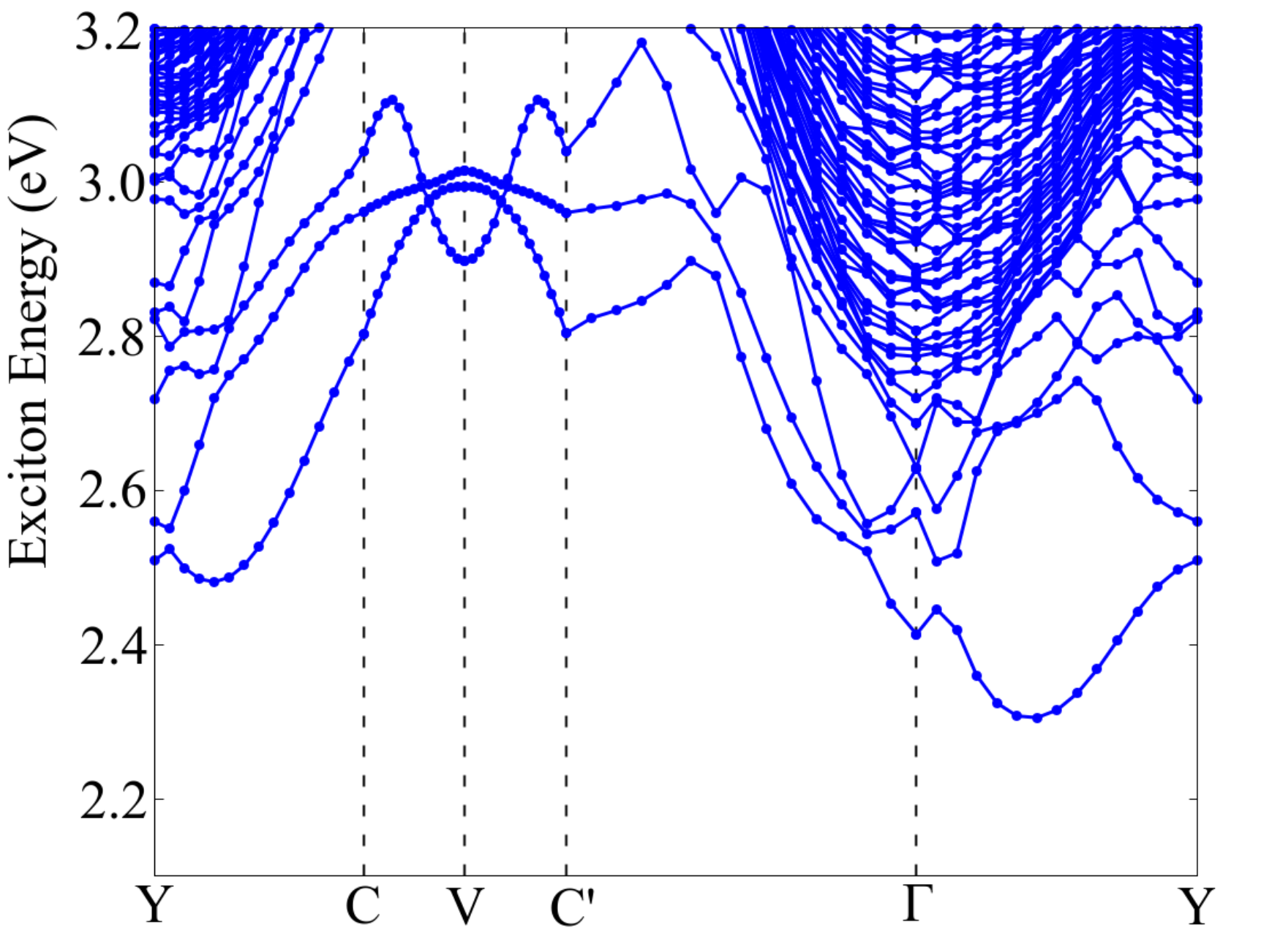}
    \caption{The excitonic band structure (DFT-HSE06-BSE) of the Tetragraphene monolayer.}\label{fig:exciton_bands}
\end{figure}

As extensively reported, excitonic effects tend to be significant in 2D materials \cite{Dias_3265_2021,Silveira_1932_2022,Moujaes_111573_2023} considering the description of the optical absorption spectrum. From the TG's excitonic band structure in Fig.~\ref{fig:exciton_bands}, we can observe that despite the direct nature of the electronic band structure, we have an indirect exciton ground state of \SI{2.30}{\electronvolt}, between the $\Gamma$ and Y symmetry points, resulting in an excitonic binding energy of \SI{0.36}{\electronvolt}. 

This nontrivial, indirect excitonic behavior can be explained by the electron-hole pair orbital characteristics combined with the Coulomb interaction that binds both, which also depends on their momentum. An analogous behavior, where the bandgap is direct and the excitonic ground state is indirect, has been reported in the literature.\cite{Santos_2044_2023,Santos_1089_2023} The direct exciton ground state is bright,i.e, optically active, with an energy of \SI{2.41}{\electronvolt}. The indirect nature of the fundamental excitonic ground state enables phonon-assisted optical transitions with photon excitation energy lower than the optical band gap of \SI{2.41}{\electronvolt}. However, since the difference between the direct and indirect excitonic ground states is low ($\approx$ \SI{0.09}{\electronvolt}), other optical band gap energy variations are not expected for phonon-assisted transitions. 


\begin{figure*}[!h]
    \centering
    \includegraphics[width=1.1\linewidth]{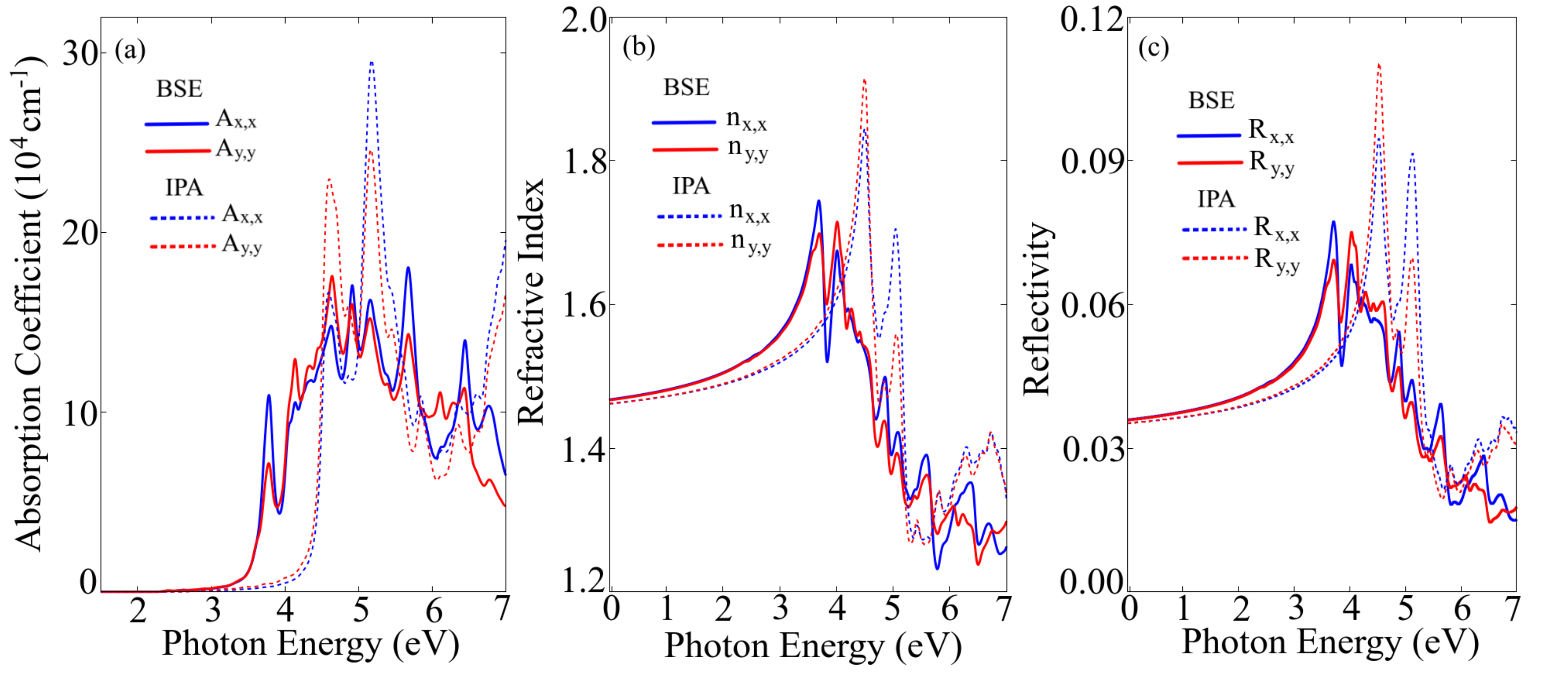}
    \caption{TG monolayer optical properties: (a) absorption coefficient, (b) refractive index, (c) reflectivity; at BSE (solid curves) and IPA (dashed curves) levels, considering linear  $\hat{x}$ (blue curves) and  $\hat{y}$ (red curves) polarizations. }\label{fig:opt-linear}
\end{figure*}

The optical linear response, at IPA and BSE approximations, represented by the absorption coefficient, refractive index, and reflectivity, are outlined in Fig.~\ref{fig:opt-linear} (a), (b), and (c), respectively. From Fig.~\ref{fig:opt-linear}(a), we can observe that the quasi-particle effects not only result in a redshift of the spectrum but also change the absorption intensity of the peaks, which are smaller at the BSE level. The system shows a small optical anisotropy independent of BSE or IPA levels, resulting in a different peak absorption intensity for $\hat{x}$ and $\hat{y}$ light polarizations. Furthermore, the optical bandgap from BSE (\SI{2.41}{\electronvolt}) and IPA (\SI{2.66}{\electronvolt}) calculations remain the same. 

Fig.~\ref{fig:opt-linear}(b) shows that at \SI{0}{\electronvolt}, the refractive index at the IPA and BSE levels is approximately $1.47$. The quasi-particle effects seem to enhance (reduce) the refractive index for optical excitations lower (higher) than \SI{4.0}{\electronvolt}. A similar comportment is shown in the reflectivity (Fig.~\ref{fig:opt-linear}(b)), which achieves a maximum value of $0.07$ ($0.09$) at the BSE (IPA) level, meaning that less than \SI{10}{\percent} of the light is reflected off this material. 

\section{Conclusions}

In this work, we have carried out a complete investigation into the mechanical, thermoelectric, and excitonic properties of a tetragraphene monolayer using DFT simulations and semiempirical methods. This unique material is characterized by a quasi-2D structure with a combination of squared and hexagonal rings built in a buckled shape.
Our findings identify the tetragraphene monolayer as a direct bandgap semiconductor with a bandgap value of about \SI{2.66}{\electronvolt} at the HSE06 level. It also exhibits an indirect excitonic ground state with an energy of \SI{2.30}{\electronvolt}, leading to an excitonic binding energy of \SI{0.36}{\electronvolt}. Furthermore, the lattice thermal conductivity of tetragraphene is approximately $\kappa_L(300)=118$ \si{\watt/\milli\kelvin}. This value is significantly lower than that of graphene ($3084.6$ \si{\watt/\milli\kelvin}), graphenylene ($600.0$ \si{\watt/\milli\kelvin}), and T-Graphene ($800.0$ \si{\watt/\milli\kelvin}), but comparable to the thermal conductivity of the biphenylene network ($208.0$ \si{\watt/\milli\kelvin}). Monolayer tetragraphene is mechanically stable, possessing an ultralow Poisson ratio of 0.09. Regarding the thermoelectric properties, our results show that the maximum figure of merit (efficiency) occurs for holes at an energy of \SI{-1}{\electronvolt} below the Fermi level. The same pattern is preserved at higher temperatures but with larger values of ZT reaching 0.04 at T=\SI{800}{\kelvin} in the case of hole doping.

\begin{acknowledgement}
The authors are grateful for the computational resources provided by Centro Nacional de Processamento de Alto Desempenho in São Paulo-CENAPAD-SP (proj 634, 897, and 909) and the Lobo Carneiro HPC (NACAD) at the Federal University of Rio de Janeiro (UFRJ) (proj 133 and 135). A.C.D. also thanks the financial support from National Council for Scientific and Technological Development (CNPq, grant number $408144/2022-0$) and Federal District Research Support Foundation (FAPDF, grants number $00193-00001817/2023-43$ and $00193-00002073/2023-84$). L.A.R.J. acknowledges the financial support from FAP-DF grants $00193.00001808/2022-71$ and $00193-00001857/2023-95$ and FAPDF-PRONEM grant $00193.00001247/2021-20$, and CNPq grant $350176/2022-1$. RMT and DSG acknowledge support from CNPq and the Center for Computational Engineering and Sciences at Unicamp, FAPESP/CEPID Grant ($2013/08293-7$).

\end{acknowledgement}

\bibliography{zboxref/ref}

\end{document}